\documentstyle[preprint,aps,eqsecnum]{revtex}
\input epsf
\begin{document}
\draft
%\preprint{dvi file made on \today}
\title
{Motion of heavy particles coupled to fermionic and bosonic
environments in one dimension}
\author{A.~O.~Caldeira $^{1,*}$ and A.~H.~Castro Neto $^{2}$}
\bigskip
\address{ $^{1}$
Department of Physics,
University of Illinois at Urbana-Champaign\\
1110 W.Green St., Urbana, IL, 61801-3080}

\address
{$^{2}$
Institute for Theoretical Physics\\
University of California\\
Santa Barbara, CA, 93106-4030}

\maketitle

\bigskip

\begin{abstract}
Making use of a simple unitary transformation we change the
hamiltonian of a particle coupled
to an one dimensional gas of bosons or fermions to a new form from
which the many body degrees of freedom can be easily traced out.
The effective dynamics of the particle allows us to compute its
damping constant in terms of the reflection coefficient of the interaction
potential and the occupation number of the
environmental particles. We apply our results to a delta repulsive
potential.
\end{abstract}

\bigskip


\narrowtext

\section{Introduction}

When one wants to study the quantum dynamics of a non-isolated system
a possible way to attack the problem is to assume that the system we
want to study is coupled to a bath of harmonic oscillators with a specific
spectral function \cite{caldeira1,caldeira2}. When this is done we are not
interested in the microscopic details of the interaction of the system
with its environment but rather in those essential features that help
us to recover the phenomenological equations of motion in the classical
limit. This has been the approach used to study dissipative effects
in the phenomena of macroscopic quantum tunneling \cite{caldeira2}
and coherence \cite{leggett}.

A second approach to investigate the quantum dynamics of a system interacting
with its environment would be  to treat the problem from a microscopic
point of view. In this case there are no standard expressions for the effective
dynamics of the system of interest and for each problem they must be
developed from the microscopic equations according to the appropriate
approximations for the particular interaction we are dealing with. There
are very important problems which are suitable for the microscopic approach,
such as; motion of ions in normal\cite{prokofev,helium}
and superfluid $He^3$ \cite{helium},
mobility of particles in metals
\cite{kondo,yamada,per} and
diffusion of heavy particles in solids \cite{clawson}, to name just a few.

Actually we could also incorporate in the microscopic approach the recently
developed formulation for dealing with mobility and diffusion of quantum
solitons in problems of coupled \cite{neto1} or non-linear field theories
\cite{neto2}. Here, the collective coordinate method \cite{rajaraman}
is shown to lead us to another very simple model \cite{neto3} for describing
quantum dissipation.

In this paper we have shown that the hamiltonian of a particle coupled
to bosons and fermions in one dimension
can also be mapped into the model of ref. \cite{neto3}. In this case
it is simply done through the application of a unitary operator which
uniformly translates the positions of all the environmental particles
changing the origin of their coordinate frame to the position of the
external particle. As a matter of fact this procedure is valid for
any interaction potential which depends only on the distance between
the particle of interest and each of the environmental particles.
At the end we specialize on the delta repulsive potential because of
mathematical convenience and transparence to analyze the effects of
foward and backward scattering processes.

We shall start by applying the above-mentioned unitary transformation
to the original hamiltonian and showing how its transformed version
can be used in the well-known Feynman-Vernon approach \cite{caldeira1,feynman}
for dealing with the reduced density operator of the external
particle. This is done in section II.

In section III we briefly review the coherent state representation
for the density operator of the system and obtain the real time
reduced density operator of the particle from which we can get its
damping constant. Although this has been presented in another paper
\cite{neto1} we decided to include it here to make this contribution
more self-contained and to correct some of our previous expressions.

In section IV we analyze the expression for the damping constant and
obtain its temperature dependence in the case of bosonic and fermionic
environments and finally present our conclusions in section V. We
have also included an appendix where we discuss the approach in
thermal equilibrium and the approximations made in our calculations.

\section{The canonical transformation}

Let us start by describing the model that will be used throughout
this paper. We assume that a particle of mass $M$ is interacting
with an one dimensional gas of $N$ fermions or bosons of mass $m$
through the hamiltonian
\begin{equation}
H = \frac{p^2}{2 M} + \sum_{i=1}^{N} U(q-q_i) + \sum_{i=1}^{N}
\frac{p_i^2}{2 m}
\end{equation}
where $p$, $q$, $p_i$, and $q_i$ are respectively the momenta and positions
of the heavy particle and of the $i^{th}$ environmental particle
 and $U(q-q_i)$ can, in principle, be any
interaction potential. We shall only assume it has a delta repulsive
form later on in this work.

The dynamics of the particle is governed by its reduced density operator
which is given by \cite{caldeira1,feynman}
\begin{equation}
\tilde{\rho}(x,y,t) = \int dq_1 dq_2 ... dq_N
\langle x, q_1,...,q_N | e^{i \frac{H t}{\hbar}} \rho(0)
e^{-i \frac{H t}{\hbar}} |y, q_1,...,q_N \rangle
\end{equation}
where $\rho(0)$ is the initial density operator of the whole system.

Now, introducing the unitary operator $U$
\begin{equation}
U = e^{-\frac{i}{\hbar} \sum_{j=1}^{N} p_j q}
\end{equation}
we can use the identity $U^{-1}U =1$ between any two operators in
(2.2) to get
\begin{equation}
\tilde{\rho}(x,y,t) = \int dq_1 dq_2 ... dq_N
\langle x, q_1,...,q_N |U^{-1} e^{i \frac{H' t}{\hbar}} \rho'(0)
e^{-i \frac{H' t}{\hbar}}U|y, q_1,...,q_N \rangle
\end{equation}
where $H'$ is the transformed hamiltonian \cite{hakim}
\begin{equation}
H' = UHU^{-1} = \frac{1}{2 M} \left(p- \sum_{j=1}^{N} p_j\right)^2
+ \sum_{j=1}^{N} \left(\frac{p_j^2}{2 m} + U(q_j)\right)
\end{equation}
and $\rho'(0) =U \rho(0) U^{-1}$  is the transformed initial condition
which we analyze below.

It should be noticed that the application of the unitary transformation
$U$ to the hamiltonian $H$ has modified it in such a way that now the
external particle interacts via momentum-momentum coupling with a set
of particles which are scattered by a potential $U(q)$. The same transformation
was used by Castella and Zotos \cite{castella} for the exact diagonalization
of the problem via the Bethe ansatz.

Now, inserting the closure relation
\begin{eqnarray}
\int dx dq_1 dq_2 ... dq_N |x, q_1,..., q_N\rangle \langle x, q_1,..., q_N|
= 1
\nonumber
\end{eqnarray}
twice between the unitary operators $U$ and $U^{-1}$ and the transformed
density operator $\rho'(t)$ one has
\begin{eqnarray}
\tilde{\rho}(x,y,t) = \int dq_1 dq_2 ... dq_N dr_1 dr_2 ... dr_N
\, \langle r_1,...,r_N|e^{-\frac{i}{\hbar} \sum_{j=1}^{N} p_j (x-y)}
|q_1,..., q_N \rangle
\nonumber
\\
\langle x,q_1,..., q_N| \rho'(t)|y,r_1,...,r_N\rangle
\end{eqnarray}
which is more cumbersome than (2.2) due to the existence of the first
term on its r.h.s.. This complication was brought about by the unitary
transformation and reflects the fact that after solving the dynamics in
the more appropriate representation we need to return to the original
representation from where we started.

Nevertheless things are not so awkward as they seem. Since we are
most of the time, and particularly in this work, interested in the average
values of observables such as $\langle x(t) \rangle$, $\langle p(t) \rangle$,
$\langle x^2(t) \rangle$ or $\langle p^2(t) \rangle$ we do not really need
(2.6) as it stands but rather its diagonal elements only.

The justification for this assertion for observables which are only
x-dependent is trivial because
\begin{equation}
\langle x^n(t) \rangle = \int dx x^n \tilde{\rho}(x,x,t)
\end{equation}
and the first term on the r.h.s. of (2.6) becomes $\delta(q_1-r_1) ...
\delta(q_N-r_N)$ which can be trivially integrated yielding
\begin{equation}
\langle x^n(t) \rangle = \int dx dq_1...dq_N x^n
\langle x,q_1,..., q_N| \rho'(t)|x,q_1,...,q_N\rangle.
\end{equation}

For p-dependent operators the demonstration is also easy although not
immediate as in (2.7-2.8). Now, what we need is
\begin{equation}
\langle p(t) \rangle = Lim_{x \to y} \frac{\hbar}{2 i}
\left(\frac{d}{d x} - \frac{d}{d y}\right) \tilde{\rho}(x,y,t)
\end{equation}
which upon the application of the derivatives to (2.6) yields
\begin{eqnarray}
\langle p(t) \rangle = \frac{\hbar}{2 i}  Lim_{x \to y}
\int dq_1 dq_2 ... dq_N dr_1 dr_2 ... dr_N
\langle r_1,...,r_N|e^{-\frac{i}{\hbar} \sum_{j=1}^{N} p_j (x-y)}
|q_1,..., q_N \rangle
\nonumber
\\
\left( \frac{d}{d x}
- \frac{d}{d y}\right)
\langle x,q_1,..., q_N| \rho'(t)|y,r_1,...,r_N\rangle .
\end{eqnarray}
In the limit $x \to y$ the first term of the integrand above becomes
once again $\delta(q_1-r_1) ...\delta(q_N-r_N)$ which can be integrated
to give
\begin{equation}
\langle p(t) \rangle = \frac{\hbar}{2 i} Lim_{x \to y}
\int dq_1 dq_2 ... dq_N  \left( \frac{d}{d x}
- \frac{d}{d y}\right)
\langle x,q_1,..., q_N| \rho'(t)|y,q_1,...,q_N\rangle
\end{equation}
and so, what we really need in both cases is,
\begin{equation}
\tilde{\rho}'(x,y,t) = \int dq_1 dq_2 ... dq_N
\langle x,q_1,..., q_N| \rho'(t)|y,q_1,...,q_N\rangle.
\end{equation}

Here one should notice that this result is valid only if we want to
compute the above-mentioned averages. Were we interested in the time
evolution of $\tilde{\rho}(x,y,t)$ this statement would not be valid.

Before leaving this section, let us try to clarify the choice we will
make for the initial condition $\rho'(0)$ in this work. As we have seen
from (2.4) $\rho'(0)$ can be prepared in terms of the eigenstates of
the transformed hamiltonian (2.5). Our particular choice will be to
assume that the particle is in a pure state (a wave packet centered at
the origin) and the
environment is in equilibrium in the presence of the particle. This
means we can write $\rho'(0)$ as
\begin{equation}
\rho'(0) = \rho_s \rho_{eq}' = \rho_s e^{- \beta H_e}
\end{equation}
where $\rho_s$ refers only to the particle and $H_e$ is given by
\begin{equation}
H_e = \sum_{j=1}^{N} \left(\frac{p_j^2}{2 m} + U(q_j)\right)
\end{equation}
Notice that this is only valid as long as one neglects the spread of the
initial wave packet about the origin when preparing the initial state of
the composite system.
\section{The Feynman-Vernon approach}

Having specified the central object of this paper we
should now choose the most appropriate representation in which
we should write it. Since our environment is composed of indistinguishable
particles (bosons or fermions) it is convenient to write the transformed
hamiltonian (2.5) in its second quantized version which reads \cite{fetter}
\begin{equation}
H' = \frac{1}{2 M} \left(p-\sum_{i,j} \hbar g_{ij} a^{\dag}_i a_j \right)^2
+ \sum_i (\hbar \Omega_i - \mu) a^{\dag}_i a_i
\end{equation}
where  $g_{ij} = \frac{1}{\hbar} \langle i | p' |j \rangle$
is the matrix element of the
momentum operator of a single environment particle between eingenstates
$|i\rangle$ and $|j \rangle$ of $H_e$ in (2.14) , $\Omega_i$ are the
eigenfrequencies of these states and $\mu$ is the chemical potential.
The operators $a_i$ and $a^{\dag}_i$
are the standard annihilation and creation operators for bosons or fermions.

Then we can use the initial conditions (2.13) in (2.12) to write the
standard Feynman-Vernon expression \cite{caldeira1,feynman}
\begin{equation}
\tilde{\rho}'(x,y,t) = \int dx' \int dy' {\cal J}(x,y,t;x',y',0)
\rho_s(x',y',0)
\end{equation}
where
\begin{eqnarray}
{\cal J}(x,y,t;x',y',0) = \int dq_1...dq_N dr_1...dr_N ds_1..ds_N
\langle x, q_1,...,q_N|e^{-\frac{i}{\hbar} H' t}|x',r_1,...,r_N\rangle
\nonumber
\\
\langle r_1,...,r_N|e^{-\beta H_e}|s_1,...,s_N\rangle
\langle y',s_1,...,s_N|e^{\frac{i}{\hbar} H' t}|y,q_1,...,q_N\rangle.
\end{eqnarray}

However, as we have chosen to represent our hamiltonian $H'$ in its
second quantized form we had better perform all the integrals in (3.3)
in the coherent state representation for the creation and annihilation
operators \cite{negele} involved therein. In doing so we can rewrite
${\cal J}$ in (3.3) as
\begin{eqnarray}
{\cal J}(x,y,t;x',y',0) = \int \frac{d^2 \vec{\alpha}}{\pi}
\frac{d^2 \vec{\gamma}}{\pi}
\frac{d^2 \vec{\delta}}{\pi}
\langle x, \vec{\alpha} |e^{-\frac{i}{\hbar} H' t}
|x',\vec{\gamma}\rangle
\langle \vec{\gamma}|e^{-\beta H_e}|\vec{\delta}\rangle
\langle y',\vec{\delta}|e^{\frac{i}{\hbar} H' t}
|y,\vec{\alpha}\rangle.
\end{eqnarray}
where $\vec{\alpha}$ is a complex vector with infinte number of components and
$\frac{d^2 \vec{\alpha}}{\pi}
= \frac{1}{\pi} \prod_i d(Re\alpha_i) d(Im\alpha_i)$.
Equation (3.4) can be written in a more appropriate form if we use the
functional integral representation for all the time evolution operators
(in real and imaginary time) of its integrand. We can write
\begin{eqnarray}
{\cal J}(x,y,t;x',y',0) = \int {\cal D}x(t') {\cal D}y(t')
e^{\frac{1}{\hbar} \left(S_0[x]-S_0[y]\right)} {\cal F}[x,y]
\end{eqnarray}
where
\begin{equation}
S_0[x] = \int_0^t dt' \frac{1}{2} M \dot{x}^2(t')
\end{equation}
is the action of the particle if it were not coupled to the environment,
${\cal F}[x,y]$, the well-known influence functional, is given by
\begin{equation}
{\cal F}[x,y] = \int \frac{d^2 \vec{\alpha}}{\pi}
\frac{d^2 \vec{\gamma}}{\pi}
\frac{d^2 \vec{\delta}}{\pi} \rho_R(\vec{\gamma}^*,\vec{\delta})
\int_{\vec{\gamma}}^{\vec{\alpha^*}} {\cal D}^2 \vec{\alpha}
\int_{\vec{\delta}^*}^{\vec{\alpha}} {\cal D}^2 \vec{\delta}
\, e^{\frac{i}{\hbar} \left(S_I[x,\vec{\alpha}]-S^*_I[y,\vec{\delta}]\right)}
\end{equation}
with
\begin{equation}
S_I[x,\vec{\alpha}] = \int_0^{t} dt' \left(\frac{i \hbar}{2}
\sum_n (\alpha^*_n \dot{\alpha}_n - \alpha_n \dot{\alpha}^*_n)
+ \dot{x} \sum_{m,n} \hbar g_{m,n} \alpha_m^* \alpha_n
- \sum_n (\hbar \Omega_n -\mu) \alpha_n^* \alpha_n \right)
\end{equation}
and
\begin{eqnarray}
\rho_R(\vec{\gamma}^*,\vec{\delta}) &=& \frac{\exp\left\{e^{-\beta (\hbar
\Omega_n-\mu)} \gamma_n^* \delta_n\right\}}{Z} ,
\nonumber
\\
Z &=& \int \frac{d^2 \vec{\alpha}}{\pi} \exp\left\{e^{-\beta (\hbar
\Omega_n-\mu)} |\alpha_n|^2\right\} .
\end{eqnarray}

The integral in (3.7) is exactly the same that has been evaluated
in the context of the polaron dynamics in ref. \cite{neto1}. The
only difference is that we now must deal with the environmental
particles as being massive bosons or fermions.
The result of the integration follows exactly as in \cite{neto1}
and reads \cite{note}
\begin{equation}
{\cal F}[x,y] = [\det(1 \mp \bar{N} \Gamma[x,y])]^{\mp 1}
\end{equation}
where the $-1$ refers to bosons and $+1$ refers to fermions,
$\bar{N}$ is the ocupation number operator given by
\begin{eqnarray}
\bar{N}_{ij} &=& \delta_{ij} \overline{n}_i
\nonumber
\\
\overline{n}_i &=& \frac{1}{e^{\beta (\Omega_i-\mu)} \mp 1}
\end{eqnarray}
and the functional $\Gamma[x,y]$ is such that
\begin{equation}
\Gamma_{nm}[x,y] = W_{nm}[x] + W^*_{nm}[y] + \sum_k W^*_{nk}[y]
W_{km}[x]
\end{equation}
with $W_{nm}$ satisfying the integral equation
\begin{equation}
W_{nm}[x,\tau] = \int_0^{\tau} dt' \, W^0_{nm}(\dot{x},t')
+ \sum_k \int_0^{\tau} dt' \int_0^{t'} dt'' \, W^0_{nk}(\dot{x},t')
W_{km}(\dot{x},t'')
\end{equation}
where
\begin{equation}
W^0_{nm}(t') = i g_{nm} \dot{x}(t') e^{i(\Omega_n-\Omega_m)t'}
(1-\delta_{nm}).
\end{equation}
So far our results are exact.

Since the influence functional ${\cal F}$ must still be integrated
in (3.5) its form (3.10) is useless. In order to evaluate
the double functional integral in (3.5) we have to make some approximations
in our present expression for ${\cal F}[x,y]$. Observing, from
(3.13) and (3.14), that $\Gamma[x,y]$ depends on the
velocities $\dot{x}$ and $\dot{y}$ we can rely on the assumption
that only slow paths will contribute to (3.5) and expand (3.10) up to quadratic
terms in $\dot{x}$ and $\dot{y}$.

This was the procedure followed in reference \cite{neto1} and therefore
we shall only quote the final result of this approximation which reads
\begin{equation}
{\cal F}[x,y] = e^{-\frac{i}{\hbar} \Phi_I[x,y]}
e^{-\frac{1}{\hbar}\Phi_R[x,y]}
\end{equation}
where
\begin{equation}
\Phi_I = \int^t_0 dt'  \; \int^{t'}_0 dt" (\dot{x}(t') - \dot{y}(t'))\hbar
\Gamma_I(t'-t") (\dot{x}(t") + \dot{y}(t"))
\end{equation}
and
\begin{equation}
\Phi_R = \int^t_0 dt' \; \int^{t'}_0 dt"
 (\dot{x}(t') - \dot{y}(t'))\hbar \Gamma_R(t'-t") (\dot{x}(t") - \dot{y}(t"))
\end{equation}
with
\begin{equation}
\Gamma_R(t) = \frac{1}{2} \sum_{i,j} |g_{ij}|^2
(\overline{n}_i + \overline{n}_j + 2\overline{n}_i\overline{n}_j)
 \cos(\Omega_i - \Omega_j)t
\end{equation}
and
\begin{equation}
\Gamma_I(t) = \sum^\infty_{i,j} |g_{ij}|^2
(\overline{n}_i -\overline{n}_{j}) \sin(\Omega_i - \Omega_j)t.
\end{equation}
{}From these expressions we can determine the damping function
$\gamma(t)$ as
\begin{equation}
\gamma(t) = -\frac{\hbar}{M} \frac{d \Gamma_I(t)}{d t} =
-\frac{\hbar}{2 M} \int d\omega \int d\omega'
S(\omega,\omega') [\overline{n}(\omega)-\overline{n}(\omega')]
(\omega-\omega') \cos(\omega-\omega')t
\end{equation}
and the diffusion function $D(t)$ as
\begin{equation}
D(t) = -\hbar \frac{d^2 \Gamma_R}{d t^2} =
\frac{\hbar^2}{2} \int d\omega \int d\omega'
S(\omega,\omega')
[\overline{n}(\omega)+\overline{n}(\omega')+2\overline{n}(\omega)
\overline{n}(\omega')] (\omega-\omega')^2
\cos(\omega-\omega')t.
\end{equation}
In the expressions for $\gamma(t)$ and $D(t)$ we have introduced
the scattering function $S(\omega,\omega')$ which we define as
\begin{equation}
S(\omega,\omega') = \sum_{i,j} |g_{ij}|^2 \delta(\omega-\Omega_i)
\delta(\omega'-\Omega_j).
\end{equation}
In this paper we shall be mostly concerned with the damping function
(3.20).

\section{The damping constant}

Having deduced the explicit form of $\gamma(t)$ our next step is
to investigate its long time behavior. As in ref. \cite{neto1}
it is our goal now to evaluate the scattering function $S(\omega,\omega')$
for a specific choice of our model hamiltonian. Let us assume that
the interaction potential between the external body and the
environmental particles is given by
\begin{equation}
U(q-q_i) = V_0 a \delta(q-q_i)
\end{equation}
where $V_0$ and $a$ are, respectively, the height and the range of
the potential barrier.

Now, because of the unitary transformation we used, the expression
of $S(\omega,\omega')$ depends only on the evaluation of the matrix
elements of the single particle momentum operator between two different
states in (2.14). As this is a hamiltonian of non-interacting particles
subject to a delta repulsive potential at the origin it is a very
simple task to obtain its eigenstates. Observe that this problem
is essentially the one of an impurity in one dimension that has been studied
using many different techniques, from bosonization and renormalization
group calculations \cite{kane,gogolin}
to Bethe ansatz approaches \cite{andreas}.

Let us suppose that our environmental particles are contained in a
very large one dimensional box of length $2 L$. It is then easy to
show that we have even and odd solutions given by \cite{lipkin},
\begin{eqnarray}
u_E(k,x) &=& \frac{1}{\sqrt{L}} \cos(k |x| + \delta_k)
\\
u_O(k,x) &=& \frac{1}{\sqrt{L}} \sin (k x)
\end{eqnarray}
where
\begin{equation}
\delta_k = \frac{1}{2} \tan^{-1} \left(\frac{Im(t+r)}{Re(t+r)}\right)
\end{equation}
with
\begin{eqnarray}
r(k) &=& \frac{i \kappa}{2 |k| D(k)}
\nonumber
\\
t(k) &=& 1 + r(k)
\nonumber
\\
D(k) &=& 1 - \frac{i \kappa}{2 |k|}
\nonumber
\\
\kappa &=& \frac{4 m V_0}{\hbar^2}.
\end{eqnarray}
Notice that the phase shifts of the odd solutions are zero and
that we can further relate $\delta_k$ to the transmission
and reflection coefficients of the delta repulsive potential
as (the case of a general potential without bound states
is treated in the Appendix)
\begin{eqnarray}
T(k) = |t(k)|^2 = \cos^2 \delta_k
\nonumber
\\
R(k) = |r(k)|^2 = \sin^2 \delta_k.
\end{eqnarray}

Our first step to get an analytic expression for
$S(\omega,\omega')$ is to evaluate
\begin{equation}
g_{kk'} = \int_{-L}^{L} \frac{dx}{L} u_E(k,x) \left(-i\frac{d}{dx}
u_O(k',x)\right)
\end{equation}
which gives
\begin{eqnarray}
g_{kk'} &=& -\frac{i}{L} \left\{ \frac{k k'}{k+k'}
\left[ \frac{\sin(k-k')L}{k-k'} \cos \delta_k -
\frac{(1-\cos(k-k')L)}{k-k'} \sin \delta_k \right] \right.
\nonumber
\\
&+& \left. \frac{k k'}{k-k'}
\left[ \frac{\sin(k+k')L}{k+k'} \cos \delta_k -
\frac{(1-\cos(k+k')L)}{k+k'} \sin \delta_k \right] \right\}.
\end{eqnarray}
When we take the limit $L \to \infty$ in (4.8) and
use that, since $g$ is purely imaginary, $g_{kk'} = - g_{k'k}$,
we have
\begin{equation}
g_{kk'} = -\frac{i}{L} \left\{ \frac{\pi}{2}
\delta(k^2-k^{'2}) k k' [\cos \delta_k -\cos \delta_{k'}]
- \frac{k k'}{2} {\cal P}\left[\frac{\sin \delta_k + \sin \delta_{k'}}{
k^2-k^{'2}} \right] \right\},
\end{equation}
with ${\cal P}$ meaning the principal value.
This expression must now be used in the continuum version of (3.22)
that reads
\begin{equation}
S(\omega,\omega') = \frac{L^2}{\pi^2} \int dk \int dk'
|g_{kk'}|^2 \delta(\omega-\Omega_k) \delta(\omega'-\Omega_{k'})
\end{equation}
where $\Omega_k = \frac{\hbar^2 k^2}{2 m}$. Evaluating (4.10) one gets
\begin{equation}
S(\omega,\omega') = \frac{m}{8 \pi^2 \hbar^2}
\sqrt{\omega \omega'} \left[\frac{\sin \delta(\omega)-\sin \delta(\omega')}
{\omega-\omega'}\right]^2.
\end{equation}

Now we can finally evaluate the integral for $\gamma(t)$ in
(3.20) which in the long time approximation gives us
\begin{equation}
\gamma(t) = \bar{\gamma}(T) \delta(t)
\end{equation}
where
\begin{equation}
\bar{\gamma}(T) = - \frac{m}{2 \pi \hbar M} \int dE \, E \, R(E) \,
\frac{d \overline{n}}{d E}
\end{equation}
is the damping constant that we want to study. In this expression the
reflection coefficient $R(E)$ can be easily obtained from (4.5) as,
\begin{equation}
R(E) = \frac{E_0}{E_0+E}
\end{equation}
where $E_0 = \frac{m V_0^2 a^2}{2 \hbar^2}$ is what we shall be calling
the strength of the delta repulsive potential.

Let us now study (4.13) for fermions and bosons at zero and finite
temperatures.

\subsection{Fermions}

At zero temperature it is straightfoward to write $\bar{\gamma}$ for the
fermionic case. As we know, at $T=0$,
\begin{equation}
 \frac{d \overline{n}}{d E} = - \delta(E - E_F)
\end{equation}
where $E_F$ stands for the Fermi energy of the system. Therefore,
\begin{equation}
\bar{\gamma}(T=0) = \frac{m}{2 \pi \hbar M} \frac{E_0 E_F}{E_0+E_F}.
\end{equation}

Equation (4.16) tell us that for strong interactions or low densities
($E_F<<E_0$) the damping
constant is dominated by $E_F$. Although the reflection coefficent tends
to one at very low energies, the number of fermions being scattered tends
to zero. On the other hand, for weak interactions or high densities
($E_F>>E_0$) the expression
for the damping constant is governed by $E_0$ because at very high energies
the reflection coefficient is vanishingly small and, consequently, all the
scattering effects come from the low energy fermions ($E \leq E_0$).

The finite temperature behavior of (4.13) will now be studied in two
limits; the low and high temperature regimes. For studying the low
temperature limit we make use of the well known Sommerfeld expansion
\cite{ashcroft} which allows us to write
\begin{equation}
\int_0^{\infty} dE f(E) \left(-\frac{d \overline{n}}{dE}\right)
\approx f(\mu) + \frac{\pi^2}{12} \frac{d^2 f(\mu)}{d \mu^2} (k_B T)^2
+ {\cal O}(T^4)
\end{equation}
where, for the one dimensional Fermi gas,
\begin{equation}
\mu \approx E_F + \frac{\pi^2}{12} \frac{(k_B T)^2}{E_F}.
\end{equation}
Using (4.14) and (4.17) in (4.13) and keeping only terms
${\cal O}(T^2)$ we have
\begin{eqnarray}
\bar{\gamma}(T) &=& \frac{m E_0}{2 \pi \hbar M} \frac{x}{1+x}
\left\{1 + \frac{\pi^2}{12} \frac{1-x}{(1+x)^2} \left(\frac{k_B
T}{E_F}\right)^2
\right\}
\nonumber
\\
x&=&\frac{E_F}{E_0}.
\end{eqnarray}

{}From (4.19) we see that the finite temperature correction to $\bar{\gamma}$
is strongly dependent on $x$. For strong interactions or low densities ($x<1$)
the low temperature correction to $\bar{\gamma}$ is positive. This means
that we are starting to populate states with $E>E_F$ in a region where
the reflection coefficient is still appreciable.

For weak potentials or high densities ($x>1$) we have already seen that
the damping is dominated by low energy states. As we increase the
temperature there is no change in the behavior of the high energy
fermions because of the very small values of the reflection coefficient
in this region and only the depletion of the low energy states will
influence the behavior of the damping constant lowering its zero
temperature value.

The classical limit, which we define as $k_B T >> E_F$, can also
be analyzed if we use that in this limit the Fermi occupation
number is the same as in the distinguishable particle case, namely
\cite{fetter};
\begin{equation}
\overline{n}(E) = \frac{\lambda}{g} \sqrt{\frac{2 \pi \hbar^2 \beta}{m}}
e^{-\beta E}
\end{equation}
where $g$ is the degeneracy factor and $\lambda$ the linear density
of fermions,
\begin{equation}
\lambda = \frac{N}{2 L} = \frac{g k_F}{\pi}.
\end{equation}

Using (4.20) and (4.13) we can approximately evaluate that integral in
two limits of interest,
\begin{eqnarray}
\bar{\gamma}(T) = \frac{m}{\pi M \hbar} \sqrt{\frac{k_B T E_F}{\pi}}
\; \; \; \; \; \; \; \;
E_0 >> k_B T
\nonumber
\\
\bar{\gamma}(T) = \frac{m}{\pi M \hbar}\sqrt{\frac{E_F^3}{\pi k_B T}}
\; \; \; \; \; \; \; \;
E_0 << k_B T .
\end{eqnarray}
Actually there are three distinct cases which
we can analyze in the whole temperature range.

a) $E_F>>E_0$; in this case $\bar{\gamma}(0)$ is proportional to $E_0$ and
starts
to decrease as the temperature is raised. It keeps on decreasing
up to high temperatures ($k_B T >> E_F >> E_0$) where it has
a $T^{-1/2}$ behavior.

b) $E_0>>E_F$; now $\bar{\gamma}(0)$ is proportional
to $E_F$ and increases with temperature. However, there must be a
maximum value for $\bar{\gamma}(T)$ since its predicted behavior
at high temperatures ($k_B T >> E_0 >> E_F$) also obeys the
$T^{-1/2}$ law.

c) Impenetrable external potential, $E_0 \to \infty$;
at low temperatures the behavior
of the damping constant is exactly the same as the previous case.
However, as the temperature increases it will never get to
the $T^{-1/2}$ behavior because $E_0>>k_B T>>E_F$. Therefore
it presents a $T^{1/2}$ for high temperatures.

The $T^{1/2}$ behavior of a) and b) above is easily understood
because in both cases the energy $k_B T$ exceeds the strength $E_0$
of the potential barrier causing a strong depletion of the states
which are sensitive to the presence of the scattering center.

For the impenetrable potential the reflection coefficient is always
one and therefore $\bar{\gamma}(T)$ is a monotonically increasing
function of $T$ and behaves as $T^{1/2}$ in the classical limit.
These temperature dependences are illustrated in Fig.(1) and Fig.(2).

\begin{figure}
\epsfysize=3.2 truein
\epsfxsize=3.2 truein
\centerline{\epsffile{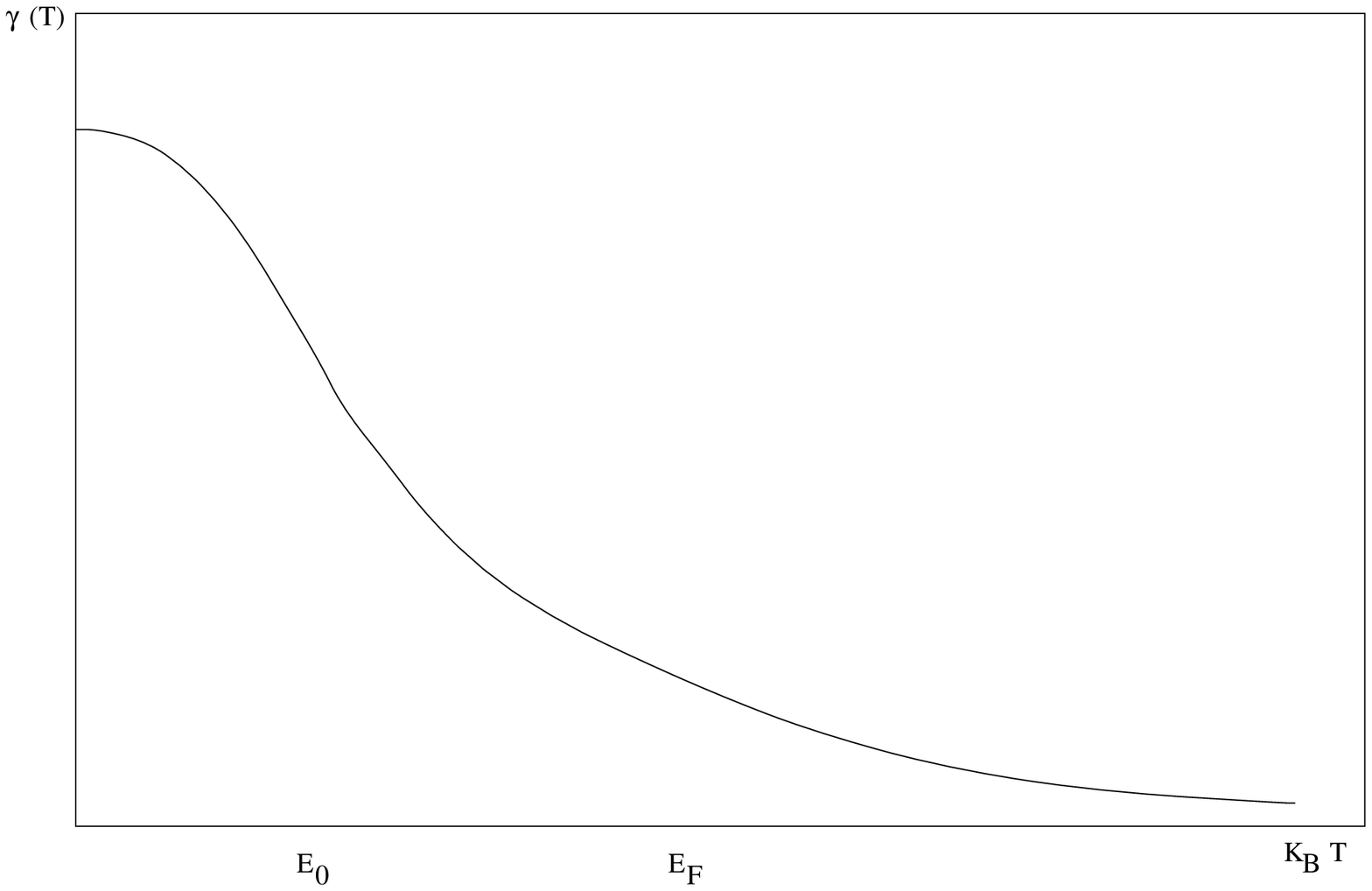}}
\caption[]{Damping constant $\gamma(T)$ as a function of temperature
for $E_0 < E_F$}
\end{figure}

\begin{figure}
\epsfysize=3.2 truein
\epsfxsize=3.2 truein
\centerline{\epsffile{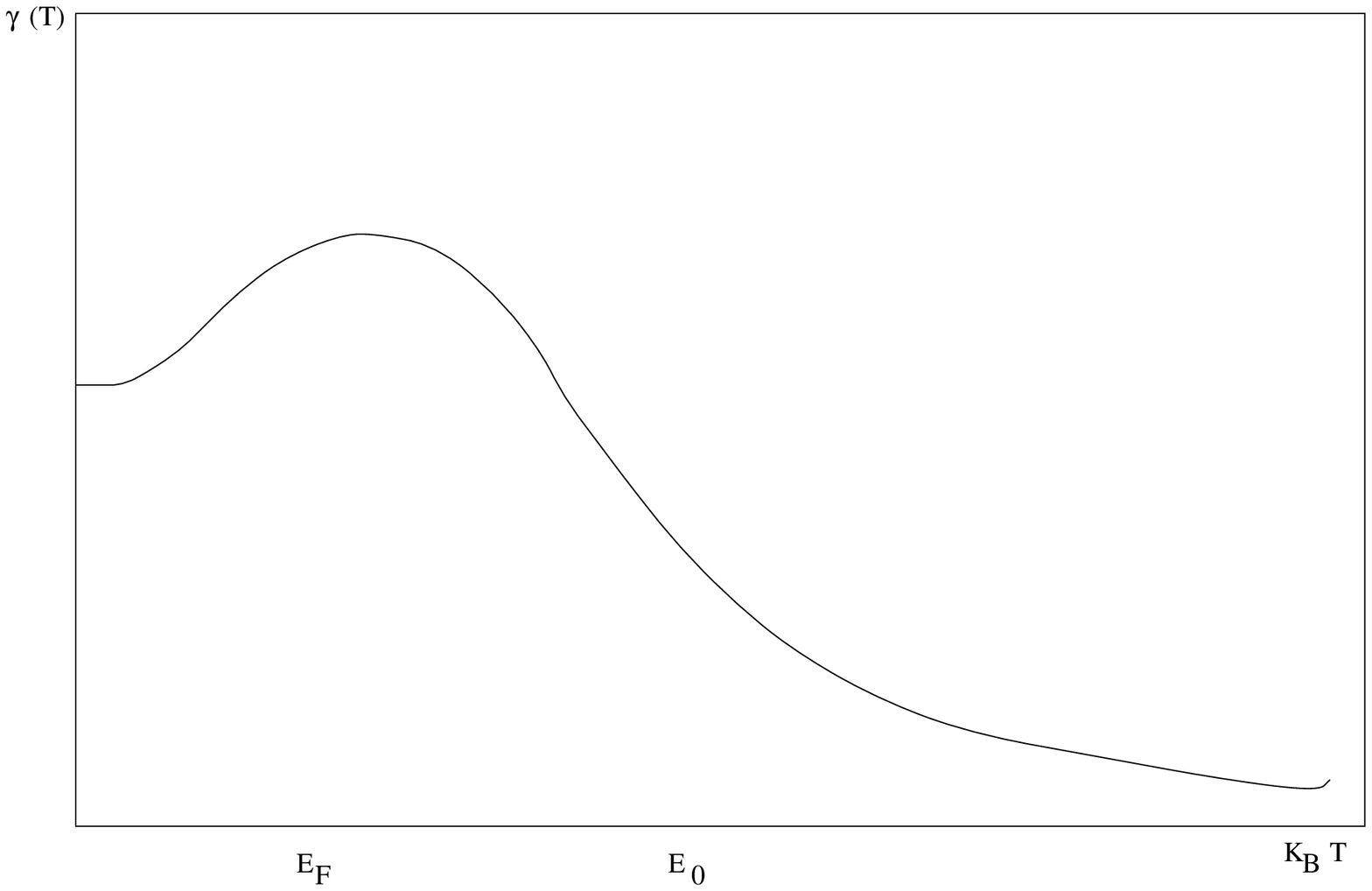}}
\caption[]{Damping constant $\gamma(T)$ as a function of temperature
for $E_0 > E_F$}
\end{figure}

\subsection{Bosons}

The same analysis can be carried on in the case of massive bosons.
The steps are not so straighforward as in the fermionic case but
they present no special difficulty either. It is a very easy task
to show that the occupation number for bosons has the
following assymptotic limits;
\begin{eqnarray}
\overline{n}(E) = \frac{k_B T}{E + |\mu|}
\; \; \; \; \; \; \; \;
E + |\mu| << k_B T
\nonumber
\\
\overline{n}(E) = e^{-\frac{E+|\mu|}{k_B T}}
\; \; \; \; \; \; \; \;
E + |\mu| >> k_B T.
\end{eqnarray}
Observe that for bosons the chemical potential $\mu$ is always
negative \cite{fetter}.

Now, if we assume that, when $T\to 0$, $\mu(T)$ behaves as \cite{dick}
\begin{equation}
\mu(T) = - \frac{k_B T}{N}
\end{equation}
we can deduce from (4.23) that
\begin{equation}
\overline{n}(E=0) = N
\end{equation}
where $N$ ist the total number of bosons in the system.

In order to further investigate the crossover from the low
to high energy behavior of $\overline{n}(E)$ in (4.23-4.24)
let us assume that we are in the high temperature regime
which means that we can replace the chemical potential
by its one dimensional classical expression \cite{fetter},
\begin{equation}
\mu_{1D}(T) \approx \frac{k_B T}{2} \ln\left(\frac{T_0}{T}\right)
\end{equation}
where $T_0 = \frac{2 \pi \hbar^2 \lambda^2}{m k_B}$ is the
temperature to localize a particle in a region of length
$2 L/N$ and $\lambda$ is, as in (4.21), the linear density
of particles. Notice that by high temperatures we mean $T > T_0$.
In this case the crossover takes place at the
energy
\begin{equation}
E_c = k_B T -|\mu| = k_B T \left[1+\frac{1}{2} \ln\left(\frac{T_0}{T}\right)
\right].
\end{equation}
Since $\ln(T_0/T)$ is a monotonically decreasing function of $T$ we see that
as we increase $T$ the crossover energy $E_c$ is pulled down to lower values
and eventually reaches zero. This means that at this temperature the
occupation number can be treated classically for any energy $E$. It is a
trivial matter to calculate this temperature because when is is reached
one has $E_c(T_c)=0$ or,
\begin{equation}
T_c \approx e^2 T_0.
\end{equation}
Therefore, this is the temperature above which the system is classical.

The same analysis tells us that when $T<<T_0$ the crossover
energy $E_c$ is very high and, consequently, the low energy form
of $\overline{n}(E)$ dominates any physical phenomenon in this
temperature range. Actually we can use (4.23) to compute
$N_0(T)$, the number of particles in the lowest energy state,
for very low $T$. We can do this by writing that the total number
of bosons can be approximated by the expression
\begin{equation}
N \approx \frac{L}{\pi} \sqrt{\frac{2 m}{\hbar^2}}
\int_0^{k_B T -|\mu|} \frac{dE}{2 \sqrt{E}} \frac{k_B T}{E+|\mu|}
= \frac{L}{\pi} \sqrt{\frac{2 m}{\hbar^2}}\frac{k_B T}{\sqrt{|\mu|}}
\tan^{-1} \sqrt{\frac{k_B T -|\mu|}{ |\mu|}}
\end{equation}
and then, assuming that
\begin{equation}
Lim_{N \to \infty, T \to 0} \frac{|\mu|}{T} = 0 ,
\end{equation}
which is in agreement with (4.24), one gets
\begin{equation}
\mu(T) \approx - \frac{k_B T}{N_0(T)}
\end{equation}
where $N_0(T) \approx \frac{4 T_0}{\pi T}$.

This is the result that enable us to compute the low temperature
behavior of $\bar{\gamma}(T)$. This is done by integrating (4.13)
by parts and using (4.23) to write
\begin{equation}
\bar{\gamma}(T) = \frac{m}{2 \pi \hbar M} \int_0^{k_B T - |\mu|} dE
\left(\frac{E_0}{E+E_0}\right)^2 \frac{k_B T}{E + |\mu|}
\end{equation}
which with the help of (4.31) yields
\begin{equation}
\bar{\gamma}(T) = \frac{m}{2 \pi \hbar M} k_B T \ln \left(\frac{4 T_0}{\pi T}
\right).
\end{equation}
{}From this formula we see that the damping constant vanishes at $T=0$
but presents a non-analytic behavior as a function of temperature in
this limit. This is due to the fact that $\overline{n}(E)$ is
proportional to $\sqrt{E} \delta(E)$ when $T \to 0$ and therefore
the weight of the states influenced by the external particle vanishes
in this limit. As the temperature is raised there is a fast increase
of this weight.

The high temperature limit of the damping constant has the same form
as for the fermionic case (as it should) with the only difference that
$E_F$ must now be replaced by $k_B T_0$. Then,
\begin{eqnarray}
\bar{\gamma}(T) = \frac{m k_B}{\pi M \hbar} \sqrt{\frac{T T_0}{\pi}}
\; \; \; \; \; \; \; \; \;
E_0 >> k_B T
\nonumber
\\
\bar{\gamma}(T) = \frac{m}{\pi M \hbar}\sqrt{\frac{(k_B T_0)^3}{\pi k_B T}}
\; \; \; \; \; \; \; \; \;
E_0 << k_B T .
\end{eqnarray}
Once again following our procedure in the fermionic case we can also
study distinct temperature dependences of the damping constant as
we show below.

a) $E_0 >> k_B T$; here $\bar{\gamma}$ vanishes at $T=0$ and
increases with temperature as in (4.33).
When $k_B T >> E_0 >> k_B T_0$ it decays
as $T^{-1/2}$ and, consequently,it must present a maximum at intermediate
temperatures ($E_0 >> k_B T >>k_B T_0$).

b) $E_0 \to \infty$; the low temperature behaviour is the same as
before but now one can never go beyond the regime $E_0 >> k_B T >>k_B T_0$
and therefore $\bar{\gamma}$ behaves like $T^{1/2}$.

Notice that we have not analyzed above the condition $k_B T_0 >> E_0$
because it does not make sense for bosons. In this case the strength
of the barrier is below the minumum energy required to localize a
boson within $2 L/N$ and so this would correspond to an undamped motion.

The interpretation of the high temperature behavior of $\bar{\gamma}$
follows similarly to what we have said for fermions.

Here we can also scketch the behavior of $\bar{\gamma}$ for a fixed
$T_0$ as we show in Fig.(3).

\begin{figure}
\epsfysize=3.2 truein
\epsfxsize=3.2 truein
\centerline{\epsffile{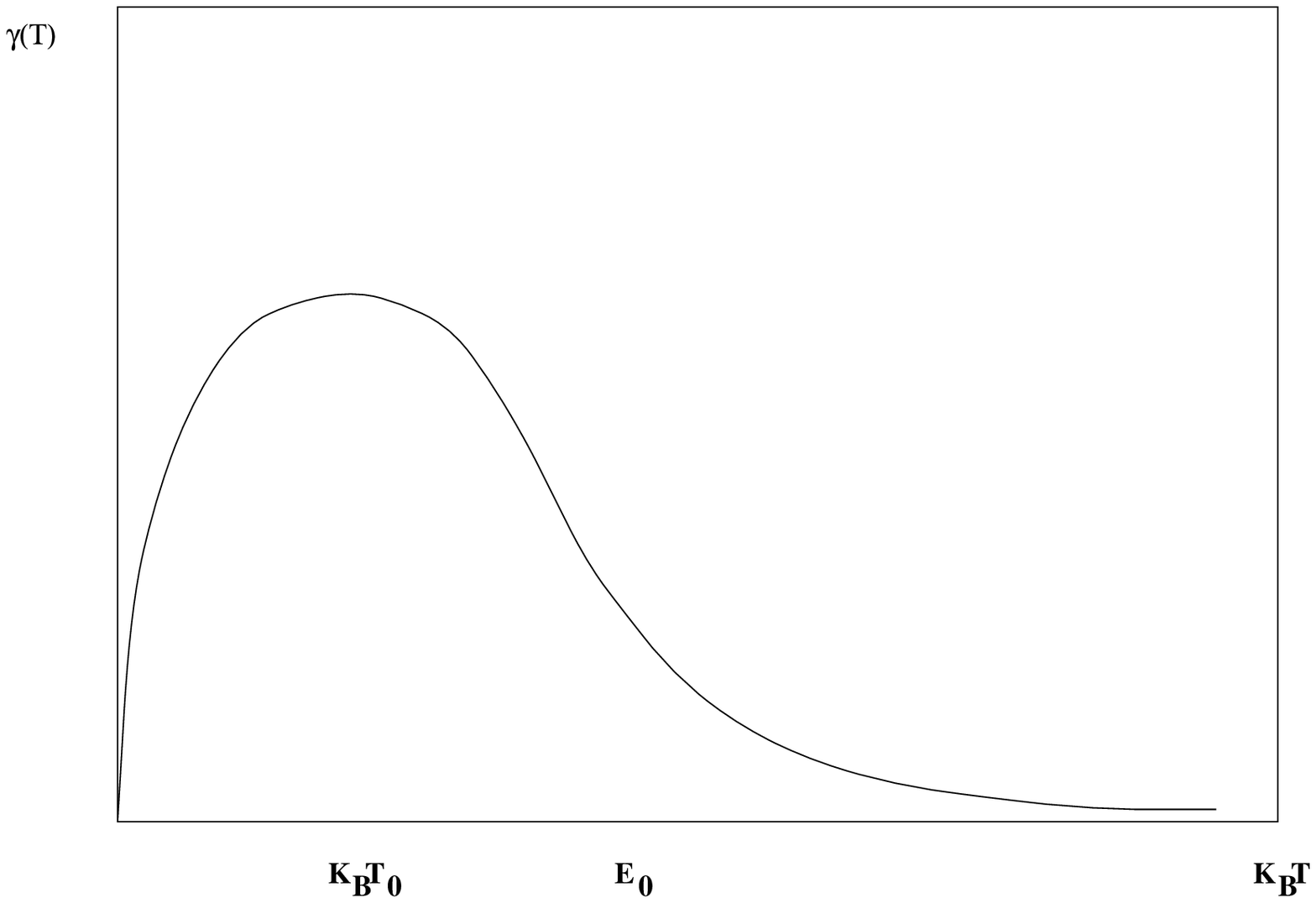}}
\caption[]{
Damping constant $\gamma(T)$ as a function of temperature
for $k_B T_0 < E_0$}
\end{figure}

\section{Conclusions}

In this paper we succeeded in writing down an expression for the damping
constant of a particle coupled via a delta repulsive potential to
non-interacting bosons and fermions.

We were able to show that using a very simple unitary transformation
we could map the original hamiltonian into a new one where the external
particle interacts via a momentum-momentum coupling to a new system
where the particles are acted by an external potential. Moreover, we
could also show that within the Feynman-Vernon picture all we needed
was to find the time evolution of the reduced density operator for
the transformed system.

Although our results were quite general very little can be done beyond
the gaussian approximation of the influence functional. On the other
hand this is exactly the form we should have got had we started with a
bath of harmonic oscillators \cite{caldeira1,caldeira2} and therefore,
a question still remains; why haven't we started with this simpler
model?

The answer is simple. In our present problem our input is the microscopic
parameters of the interaction while in \cite{caldeira1,caldeira2} all
one has is a constraint to recover the Langevin equation in the classical
limit. There, the microscopic constants all combine in order to establish
the low frequency behavior of the spectral function $J(\omega)$
\cite{caldeira1,caldeira2} to be the form $\eta \omega$, where $\eta$ is the
damping constant, whereas in the present approach the damping constant
is obtained as a function of the microscopic parameters and the
temperature. It is clear one could try to find a temperature dependent
spectral function $J(\omega,T)$ in order to recover our function
$\bar{\gamma}(T)$. However, it is hard to imagine that this could be
done a priori, without the knowledge of the specific form of
$\bar{\gamma}(T)$. Even if we can phenomenologically describe
 $\bar{\gamma}(T)$ in a vast temperature
range it is many times of great importance to
understand this behavior from first principles.

Notice that in the present formulation we have also a prescription
to go beyond the gaussian approximation. This can be used if one
wants to test the stability of the ``mean field'' solution to this
problem. This project is being carried forward by one of us
in a separate contribution \cite{mpa}.

\acknowledgments
We are deeply indebt to H.Castella, A.O.Gogolin,
M.P.A.Fisher, A.W.W.Ludwig, N.Prokofev and X. Zotos
for many useful comments and discussions.
A.O.C. is also grateful to the partial support
of the Conselho Nacional de Desenvolvimento Cient\'ifico e
Tecnol\'ogico (CNPq-Brazil). This work was partially supported by the
NSF through grants DMR92-14236 and PHY89-04035.

\newpage

\appendix

\section{Formulation in thermal equilibrium}

In this appendix we extend the real time calculation of the
paper to imaginary time and stress the approximations used
in the bulk of the paper.

We look here at the partition function for the initial
problem which can be written as,
\begin{equation}
Z = tr\left(e^{-\beta H}\right) = tr\left(e^{-\beta H'}\right)
\end{equation}
where we have used the unitary transformation (2.3) and
the fact that the trace is invariant under this transformation.

The partition function can be written in terms of path integrals
for the external particle and the fields of the bath as
\begin{equation}
Z = \int Dq \int D\Psi^* D\Psi \exp
\left(-S_E \left[q,\Psi^*,\Psi\right]\right)
\end{equation}
where the action associated with the hamiltonian (2.5) reads ($\hbar =1$),
\begin{equation}
S_E = \int_0^{\beta} dt \left\{\frac{M \dot{x}^2}{2} -i \dot{x}
\sum_{k,k'} g_{kk'} \Psi^{*}_k \Psi_{k'} - \sum_k
\left(\frac{d}{dt}-E_k\right) \Psi^{*}_k \Psi_k \right\}.
\end{equation}

Since we are interested only in the dynamics of the particle we
can trace the environmental particles exactly in order to get,
\begin{equation}
Z= \int Dq \exp\{-S_{eff}[q]\}
\end{equation}
where the effective action for the particle is non-linear and
reads
\begin{equation}
S_{eff} = \int_0^{\beta} dt \frac{M \dot{x}^2}{2}
\mp \ln \det W[q]
\end{equation}
where the minus sign is for fermions and the plus sign for bosons.
The operator $W$ is written in terms of its matrix elements as,
\begin{equation}
\langle k t |W| k' t' \rangle = \delta(t-t') \left[ \left(\frac{d}{dt}
-E_k\right) \delta_{kk'} - i  \dot{x} g_{kk'} \right].
\end{equation}

Clearly the problem (A5) is extremely complicated and is equivalent
to the one obtained in real time from expression (3.10). However
we are interested only in the case where the velocity of the particle
is small or its mass is much larger than the mass of the environmental
particles. Therefore, we look at the saddle point equations for the
action (A5) in the limit where $M \to \infty$. It is very easy to
show that within the mean field approximation the saddle point equations read
\begin{equation}
\ddot{x} \pm \frac{i}{M} \sum_{k,k'} g_{kk'} \left(\frac{d}{dt'}
G(k,t;k',t')\right)_{t'=t} = 0
\end{equation}
and
\begin{equation}
\left(\frac{d}{dt}-E_k\right) G(k,t;k',t') \mp i \dot{x}(t)
\sum_{k''} g_{kk''} G(k'',t;k',t') = \delta_{kk'} \delta(t-t')
\end{equation}
In the limit $M \to \infty$ we choose the particular solution
\begin{eqnarray}
x(t) &=& 0
\nonumber
\\
\left(\frac{d}{dt}-E_k\right) G_0(k,t;k',t') &=& \delta_{kk'} \delta(t-t')
\end{eqnarray}
which can be easily solved by
\begin{equation}
G_0(k,t;k',t') = \frac{\delta_{k,k'}}{\beta} \sum_n \frac{e^{i\omega_n
(t-t')}}{i \omega_n - E_k}
\end{equation}
where $\omega_n = \frac{2 \pi n}{\beta}$ for bosons and $\omega_n =
\frac{ \pi (2n+1)}{\beta}$ for fermions.

Now we look at gaussian fluctuations around this mean field by
expanding the determinant in (A5) as
\begin{equation}
\ln \det W =\ln \det (H_0 + H_I) \approx \ln \det G_0^{-1} + tr(G_0 H_I)
- \frac{1}{2} tr(G_0 H_I)^2
\end{equation}
where $H_0 G_0 = 1$ and $H_I$ is the second term on the r.h.s. of (A6).

Thus, the effective action reads
\begin{equation}
S_{eff} \approx S_{eff}^0 + \int_0^{\beta} \frac{M \ddot{x}^2}{2}
+\int_0^{\beta} dt \int_0^{\beta} dt' F(t-t') \dot{x}(t) \dot{x}(t')
\end{equation}
where
\begin{equation}
F(t) = \sum_{k,k'} |g_{kk'}|^2 G_0(k,t) G_0(k',-t).
\end{equation}

It is useful to consider the action (A12) in Matsubara representation
\cite{fetter}
\begin{equation}
\Delta S_{eff} = \frac{1}{\beta} \sum_n \left(M \omega_n^2 + \omega_n^2
F(\omega_n)\right) x(\omega_n) x(-\omega_n)
\end{equation}
where
\begin{equation}
F(\omega_n) = \sum_{k,k'} \frac{|g_{k,k'}|^2}{i \omega_n + E_k -E_{k'}}
\left(\bar{n}_k - \bar{n}_{k'}\right).
\end{equation}
The actual expression for $F(\omega_n)$ now depends on the form
of the coupling constant $g_{kk'}$. In the paper we considered the
case of a delta repulsive interaction but in the following subsection
we will consider the case of a general potential without bound states.
This approach generalizes the results of section IV.

\subsection{Matrix element}

Let us consider the case where the potential of interaction
between the particle and the environmental particles
 is localized, that is, $V(x) = 0$
for $x > |a|$. In this case the Schr\"odinger equation for the
impurity problem in the assymptotic limit can be written as
\begin{equation}
\frac{d^2}{d x^2} \Psi_k(x) + k^2 \Psi_k(x) = 0
\end{equation}
where $E_k = \frac{k^2}{2 m}$.

The general solution of this problem can be written in terms
of even and odd wavefunctions as follows \cite{lipkin};
\begin{eqnarray}
\Psi^E_k(x) &=& \sqrt{\frac{2}{L}} \cos\left(k |x|+ \delta_E(k)\right)
\nonumber
\\
\Psi^O_k(x) &=& \sqrt{\frac{2}{L}} sgn(x) \sin \left(k |x|+\delta_O(k)\right)
\end{eqnarray}
where $sgn(x) = +1(-1)$ if $x>0 (x<0)$, and $\delta_E(k)$ and $\delta_O(k)$
are the even and odd phase shifts, respectively.

The phase shifts are related to the reflection and transmission coefficients
by a simple relation
\begin{eqnarray}
r(k) &=& \frac{e^{2 i \delta_E(k)}-e^{2 i \delta_O(k)}}{2}
\nonumber
\\
t(k) &=& \frac{e^{2 i \delta_E(k)}+e^{2 i \delta_O(k)}}{2},
\end{eqnarray}
which can be inverted in order to give
\begin{eqnarray}
\delta_E(k) &=& \frac{1}{2} \tan^{-1} \left(\frac{Im(t(k)+r(k))}{Re(t(k)+r(k))}
\right)
\nonumber
\\
\delta_O(k) &=& \frac{1}{2} \tan^{-1} \left(\frac{Im(t(k)-r(k))}{Re(t(k)-r(k))}
\right).
\end{eqnarray}
The reflection and transmission coefficients can be now expressed in terms of
the phase shifts as
\begin{eqnarray}
R(k) = |r(k)|^2 =\sin^2\left(\delta_E(k)-\delta_O(k)\right)
\nonumber
\\
T(k)=|t(k)|^2 = \cos^2\left(\delta_E(k)-\delta_O(k)\right).
\end{eqnarray}
Thus, any information from the scattering process is encapsulated in
the phase shifts.

{}From (3.1) it is easy to see that the coupling constant $g_{kk'}$
only couples states with different parities, that is,
\begin{eqnarray}
g^{O,O}_{kk'} = g^{E,E}_{kk'}=0
\nonumber
\\
g^{O,E}_{kk'} = - g^{E,O}_{k'k}
\end{eqnarray}
The calculation of the matrix element follow the same steps as in
(4.7) and (4.8) and reads
\begin{eqnarray}
g^{O,E}_{kk'} = \frac{i}{2 L} \frac{k k'}{k+k'} {\cal P} \left(
\frac{\sin\left(\delta_O(k)-\delta_E(k')\right)
+\sin\left(\delta_O(k')-\delta_E(k)\right)}{k-k'} \right).
\end{eqnarray}

The integral we need to evaluate in (A15) can now be written as
\begin{equation}
F(\omega_n) = \frac{1}{4 L^2} \sum_{k,k'}
\frac{(\sin\left(\delta_O(k)-\delta_E(k')\right)
+\sin\left(\delta_O(k')-\delta_E(k)\right))^2
\left(\bar{n}_k-\bar{n}_{k'}\right)}{(k^2 - (k')^2)^2 (E_k-E_{k'}
+i \omega_n)}.
\end{equation}
This integral can be done by choosing a convenient countor in the
complex plane which avoids the double pole. The final result is
\begin{equation}
F(\omega_n) = \frac{1}{L |\omega_n|}
\sum_{p} p^2 R(p) \left(-\frac{d \bar{n}_p}{d p}\right).
\end{equation}
Thus, from (A24) and (A14), the effective action reads
\begin{equation}
\Delta S_{eff} = \frac{M}{\beta} \sum_n \left(\omega_n^2 + \gamma |\omega_n|
\right) x(\omega_n) x(-\omega_n)
\end{equation}
which is the usual action for the motion of a particle in a dissipative
environment \cite{caldeira2} with a damping constant
\begin{equation}
\gamma = \frac{1}{M L} \sum_p p^2 R(p) \left(-\frac{d \bar{n}_p}{d p}\right).
\end{equation}

Equation (A26) can be rewritten using the usual definition of the
density of states
\begin{equation}
D(E) = \frac{1}{L} \sum_p \delta(E-E_p)
\end{equation}
as
\begin{equation}
\gamma = \frac{m}{M} \int_{-\infty}^{\infty} dE D(E) E R(E) v(E)
\left(-\frac{d \bar{n}(E)}{d E}\right)
\end{equation}
where $p^2 = 2 m E$ and $v(E) = \frac{d E}{d p}$. It turns out that
in one dimension the density of states is simply
\begin{equation}
D(E) = \frac{1}{2 \pi v(E)}
\end{equation}
and (A30) becomes identical to (4.13).

\newpage

$*$ On leave of absence from Instituto de F\'isica ``Gleb Wataghin",
Departamento de F\'isica do Estado S\'olido e Ci\^encia dos Materiais,
Universidade Estadual de Campinas,
13083-970, Campinas, SP, Brazil.\\


\begin{references}
\bibitem{note}
Notice that expression (4.27) in ref.\cite{neto1} is not entirely
correct. However, due to the gaussian approximation performed
later in eq. (4.32) the expression for the damping function remains
unchanged whereas the diffusion constant is only slightly altered
in the high temperature limit. Their correct expressions are (3.20)
and (3.21) of the present paper.
\end{references}
\end{document}